\documentclass[aps,pra,showpacs,twoside,twocolumn,10pt]{revtex4-1}
\usepackage[colorlinks=true, citecolor=red, urlcolor=blue ]{hyperref}
\usepackage{appendix}
\usepackage{epsfig,newlfont,amssymb,amsfonts,amsmath,bm,subfigure,palatino,mathtools,amsthm,braket,times,soul,enumitem,color}
\usepackage[normalem]{ulem}

\begin{document}

\title{Quantum thermal transistors: Operation characteristics in steady state versus transient regimes}
\author{Riddhi Ghosh, Ahana Ghoshal and Ujjwal Sen}
\affiliation{Harish-Chandra Research Institute, HBNI, Chhatnag Road, Jhunsi, Allahabad - 211 019, India}

\begin{abstract}
We show that a quantum thermal transistor can also cause the transistor effect – where one out of three terminals can control the flow of heat current in the other two – with good amplification properties in the transient regime for certain paradigmatic initial states. We find three broad classes of transient quantum thermal transistors – the first having a smaller amplification than the steady state quantum thermal transistor, the second with better amplification but a smaller operating region in terms of temperature, and the third that gives higher amplification with a larger operating region. The last type is of particular interest as it also operates in the region where the steady state thermal transistors lose the transistor effect. We discuss in some detail certain initial states for which the cases of necessarily transient transistors arise. We analyze the time variation of the amplification factor of transient thermal transistors and estimate the preferable time and duration for which they can work efficiently. 
Cumulative studies of the differences in magnitudes of the amplifications of heat currents at the non-base terminals of the quantum thermal transistor, first with respect to the base-terminal temperature and next with time, are also presented. 
\end{abstract}

\maketitle
\section{Introduction}
\label{sec:intro}
With the advancement of technology and miniaturization of digital circuits, managing and controlling thermal emissions in the 
%nanoscale 
small scale
has gained utmost importance. The 
%nanoscale 
miniature
circuits are prone to damage due to even minute thermal and electrical fluctuations. Devices to control the flow of electricity have existed and been improved upon since the invention of the vacuum diode \cite{bk1,bk2,bk3}. Over the past few decades, heat conduction in such circuits has been intensively studied \cite{a2,a3,a4,a51,a5} and devices for regulating heat flow have been envisioned \cite{a6,a7,a8,a9,a91}. 
\par 
An electrical transistor \cite{a10_1,a10_2,a10_3,a10,bk1,bk2,bk3} is a three-terminal device – one of the terminals modulates (controls) the flow of electricity through the other two. The thermal counterpart of the electrical transistor – thermal transistor \cite{a11,a12,a12_1} – regulates the heat flow in a circuit. These devices can work as heat switches and modulators. Thermal transistors have been designed by utilizing gas-liquid transitions \cite{e1}, suspended graphene \cite{e2}, and many more  systems \cite{e3,e4,e5,e6}. A recent experiment has successfully implemented the thermal transistor using a thin layer of MoS\(_{2}\) that is effective even at room temperature \cite{a13}.\par 
The analysis of thermal devices at the quantum level, especially from the viewpoint of quantum information processing and quantum computers has resulted in the development of individual quantum systems \cite{a14,a15,a16} over the past few years. Such devices include photon rectifiers \cite{a17}, photon transistors \cite{a18,a19}, quantum rectifiers \cite{a17_1,a17_2}, quantum refrigerators and heat engines \cite{R1,R2,R3,R4,R5,R4_1,R6,R7,Tr1,R6p} among others. The aim of practical realization of such devices has made quantum thermodynamics indispensable to the field \cite{T1,T2,T3,T4,T4_1,T5,T6,n1}. The macroscopic laws of thermodynamics were 
%implemented 
taken over to the
%in 
quantum scale in, e.g., Refs. \cite{T7,T8,T9,T10,T12,T13,T14,T15}. Furthermore, the possibility of interaction of a quantum system with its environment has given rise to several studies of quantum systems that talk with heat baths \cite{a20,a21,a22}. \par
In recent years, there have been several proposals for various types of quantum thermal machines. %\textcolor{blue}{
A quantum thermal transistor (QTT) constructed using two-level systems is demonstrated in \cite{Tran4,Tran5} and one with coupling between a two-level and a three-level system in \cite{ex1}.
%} 
A QTT using superconducting circuits is designed in \cite{Tran8}, where this device is said to have properties similar to the  conventional semiconductor transistor. A QTT with a giant heat amplification phenomenon in the presence of a strong system-bath coupling limit by applying the polaron transformed Redfield equation combined with full counting statistics is investigated in \cite{Tran7}. \par 
In \cite{Tran4}, a quantum thermal transistor has been proposed using an arrangement of three two-level systems, each paired with a different bosonic heat bath. Dynamic control of heat flow in two terminals is achieved by controlling the temperature of the bath attached to the third terminal. A similar device is discussed in \cite{ex1}. Another such thermal machine with qubit-qutrit coupling between the systems is shown to behave as a heat current stabilizer \cite{Tran6}. See \cite{ex2,ex3,ex4,ex6,n2} for more work in this direction. Realization of a quantum thermal transistor using defects in solid-state matter, an optomechanical system, or transmon qubits has been suggested in Ref.~\cite{n2}. A quantum thermal transistor using superconducting circuits is discussed in Ref.~\cite{Tran8}. Similar setups can potentially be used to implement the system discussed in this paper.
%
%The physical systems for the realization of a quantum thermal transistor is also being studied. A QTT using either defects in solid-state matter or optomechanical system or transmon qubits have been suggested in \cite{n2} and one using superconducting circuits in \cite{Tran8}.} See \cite{ex2,ex3,ex4,ex6,n2} for more work in this direction.  
\par
A quantum thermal device operating in the steady state can take a finite amount of time to reach close to this regime. Since qubits are susceptible to environmental fluctuations, there may exist situations where such a delay could result in decoherence and loss in control of the qubits in question, thereby limiting its operating conditions. To avoid these difficulties, the operation of such devices in the transient regime is studied; in certain cases, this is more beneficial than restricting to the steady state regime. See \cite{Tr1,R5, R7} for examples of transient quantum devices. 
Here,
%In our paper, 
we have studied whether a quantum thermal transistor can function efficiently in the transient regime and what advantages
%, \textcolor{red}{if any}, 
it provides over those operating in the steady state. In particular, we identify parameter regimes where a quantum thermal transistor is necessarily transient. \par 
%\textcolor{red}{pare dekhbo!}
The rest of the paper is arranged as follows. We have discussed a model of the quantum thermal transistor and some properties that provide a good transistor effect, by comparing the operations of quantum and classical transistors in Sec. \ref{sec:tools}. The working of a transient quantum transistor is discussed in Sec. \ref{sec:transient}. Certain cases of transient transistors and their amplification properties are included in Secs. \ref{sec:ghz} and 
%Sec. 
\ref{sec:random}. The interesting phenomenon of necessarily transient quantum thermal transistors is presented in 
%most important result of our observations are given in 
Sec. \ref{sec:ntrans}.
%:
%where it is clearly evident that 
%for some initial states, transient transistors provide substantial advantage over the steady state quantum thermal transistor; we can thus call these quantum thermal transistors as being necessarily transient. 
In 
these 
%all the above 
sections, we discuss the variation of the dynamic amplification factor with temperature of the bath connected to the “base” qubit. Section \ref{sec:time_vary}  is about the change of amplification factor with time. Finally, in Sec. \ref{sec:diff_alpha}, we discuss the difference of magnitudes of the two amplification factors with respect to temperature of the base-qubit bath for fixed time and with times for fixed temperature of all the baths. A conclusion is presented in Sec. \ref{arati-mukhopadhyay}.
\section{Underlying tools}
\label{sec:tools}
A quantum thermal transistor can be structured by three two-level systems (TLS) interacting with each other and each system connected to a thermal bath. This system was considered in the steady state regime in 
%We have followed the set-up same as the one in 
\cite{Tran4}, while we consider it at a general point of time, including transient regimes.
%where the QTT is operating in steady state regime. 
We consider a combined system consisting of three TLS – labeled as A, B, and C; the energy difference between the two levels for each TLS is taken to be zero. The temperatures of the three thermal baths are $\tilde{T}_{A}$, $\tilde{T}_{B}$, and $\tilde{T}_{C}$, respectively. The coupling constants between the TLS are proportional to $\omega_{AB}$, $\omega_{BC}$, and $\omega_{CA}$. These three TLS are analogous to the three terminals (emitter, base, and collector) of a classical electronic transistor. For our purpose, the TLS are a system of interacting spins, with only two possible states – up $(\ket{0})$ and down $(\ket{1})$, which are the eigenstates of Pauli matrix $\sigma_{z}$. 
%In this case, t
The Hamiltonian of the three TLS is given by 
\begin{center}
\begin{equation}
H_{sys} =\sum
%_{\substack{X,Y=A,B,C\\X\neq Y}} 
\frac{\hbar \omega_{XY}}{2} \sigma_{z}^{X} \sigma_{z}^{Y},
\end{equation}
\end{center}
where the sum is over the ordered pairs \((X, Y) \in \{(A,B), (B, C), (C, A)\}\). 
The state of the system is supported on the space spanned by 
%can be in either of 
the eight possible eigenstates of $H_{sys}$, viz., $\ket{1}=\ket{000}$, $\ket{2}=\ket{001}$, 
$\ket{3}=\ket{010}$, $\ket{4}=\ket{011}$, $\ket{5}=\ket{100}$, $\ket{6}=\ket{101}$, $\ket{7}=\ket{110}$ and $\ket{8}=\ket{111}$; the corresponding eigenenergies are $E_1,\,E_2,\,\ldots,\,E_8$ respectively. Although the interaction between the particles of the system is of the \(zz\) type only, it is effectively 
quantum mechanical. This is first of all because the initial states are not always products over states magnetized in the \(z\) direction. Also, the system-bath interaction Hamiltonian (see below) couples the bath creation and annihilation operators with the system's transverse magnetization operator \(\sigma_x\). In effect, these two facts render the system's dynamics being driven by an effective transverse Ising model; the transverse nature lingers even as the system evolves. Let us also mention here that the “currents” considered in this paper correspond to heat exchange 
(entropy exchange) 
and not an electronic current as in a regular transistor.
The local Hamiltonian of the three bosonic baths is given by 
\begin{equation}
 H_{bath}=\sum_{k}\hbar\omega_{k}\hat{a}_{k}^{X \dagger}\hat{a}_{k}^{X}   
\end{equation}
and the system-bath interaction Hamiltonian has the form -
\begin{equation}
H_{sys-bath}=\sigma_{x}^{X}\sum_{k} \hbar g_{k}(\hat{a}_{k}^{X}+\hat{a}_{k}^{X \dagger})    
\end{equation}
%\textcolor{blue}{
Considering the rotating wave approximation, we have only taken the energy conserving terms.
%}
Here, \(\hat{a}_{k}^{X} \ (\hat{a}_{k}^{X \dagger})\)
%a_{k}^X (a_{k}^{\dagger})$ 
is the bosonic annihilation (creation) operator, 
%$k=1,2,3$ 
and $X=A,B,C$. This would mean that the valid transitions are the ones where only one spin is flipped. Thus there are only 12 possible transitions, as discussed in \cite{Tran4}. The system-bath coupling parameters \(\omega_{AB}\) and \(\omega_{BC}\) are assumed to be equal (and \(=\delta\)), and the third one, $\omega_{CA}$, is taken to be zero. So, the eigenenergies of $H_{sys}$ reduce to $E_1=\delta\hbar, E_2=0, E_3=-\delta\hbar, E_4=0, E_5=0,
E_6=-\delta\hbar, E_7=0, E_8=\delta\hbar$. The allowed transitions for bath $A$ are $1\rightleftarrows 5,\,2\rightleftarrows 6,\,3\rightleftarrows 7,\,4\rightleftarrows 8$, for bath $B$ are $1\rightleftarrows 3,\,2\rightleftarrows 4,\,5\rightleftarrows 7,\,6\rightleftarrows 8$, and for bath $C$ are $1\rightleftarrows 2,\,3\rightleftarrows 4,\,5\rightleftarrows 6,\,7\rightleftarrows 8$, and the %\textcolor{blue}{
corresponding transition energies are $\hbar\omega_{ij}$ for the transition $i\rightarrow j$, where $\hbar\omega_{ij}=E_i-E_j$ and $\omega_{ij}=-\omega_{ji}$.
%}
%$T_B=\frac{T_B}{\delta}$ is a dimensionless quantity. 
\par In the Born-Markov approximation, the master equation of the system obeys the dynamic evolution, 
%\textcolor{blue}{
\begin{center}
\begin{equation}
\frac{d\rho}{d\tilde{t}}\ = \ -\frac{i}{\hbar}[H_{sys},\rho]\ + \ \sum_{X=A,B,C}\mathcal{L}_{X}\ [\rho],
\label{eq:dynamics}
\end{equation}
\end{center}
with $\tilde{t}$ representing time. For all further discussions, we have rescaled  time as  $\delta \tilde{t}$, and used the symbol \(t\) for it.
%}. 
$\rho$ is the density matrix of the composite system
%$t$ is the time 
and  the Lindblad operators $\mathcal{L}_{X}[\rho]$ can be written as in \cite{a22}:
%\textcolor{blue}{
\begin{center}
\begin{eqnarray}
\nonumber
\mathcal{L}_{X}\ [\rho]\ = \ \quad \quad \quad \phantom{jotto sab golmal sala mairi amar khetrei}
%btw, eTa ami nijekei galagal dich-chhi!!
\\ \nonumber \sum_{\omega \geq 0}\mathcal{J}(\omega) (1+n_{\omega}^{X}) [A_{X}(\omega) \rho A_{X}^{\dagger}(\omega) - \frac{1}{2} \{\rho, A_{X}^{\dagger}(\omega) A_{X}(\omega)\}]\\
+ \mathcal{J}(\omega)n_{\omega}^{X}[A_{X}^{\dagger} (\omega) \rho A_{X} (\omega) - \frac{1}{2} \{\rho, A_{X}(\omega) A_{X}^{\dagger} (\omega)\}]. \quad \quad
\end{eqnarray}
\end{center}
Here, $n_{\omega}^{X}=\frac{1}{e^{\hbar\omega/ K_{B}\tilde{T}_{X}}-1}$ is the Bose-Einstein distribution and $\mathcal{J}(\omega)$ is the Ohmic spectral function which is taken to be $\mathcal{J}(\omega)=\frac{g_{\omega}^2}{\omega_0}$, where $\omega_0$ is 
%an arbitraty 
a
constant having the unit of frequency. Here we have taken $g_\omega=\kappa\omega$, where $\kappa$ is a dimensionless constant.
%} 
%as in \cite{Tran4}. 
The Lindblad operators, which are the decomposed form of the system operators $(\sigma_{x}^X)$ in the nondegenerate eigenbasis of $H_{sys}$,
are given by 
%\textcolor{blue}{
\begin{equation}
A_{X}(\omega)=\sum_{\substack{i,j\\j>i\\\omega=\omega_{ij}}}\ket{i}\bra{i}\sigma_{x}^{X}\ket{j}\bra{j},
\end{equation}
%}
where $\ket{i}$ and $\ket{j}$ are elements of the nondegenerate eigenbasis of $H_{sys}$ and $A_{X}(\omega)$ is an eigenoperator of $H_{sys}$ corresponding to the energy eigenvalue $-\hbar\omega$, i.e., $[H_{sys},A_{X}(\omega)]=-\hbar\omega A_{X}(\omega)$.
\par
\begin{figure}[h]
\includegraphics[width=\linewidth]{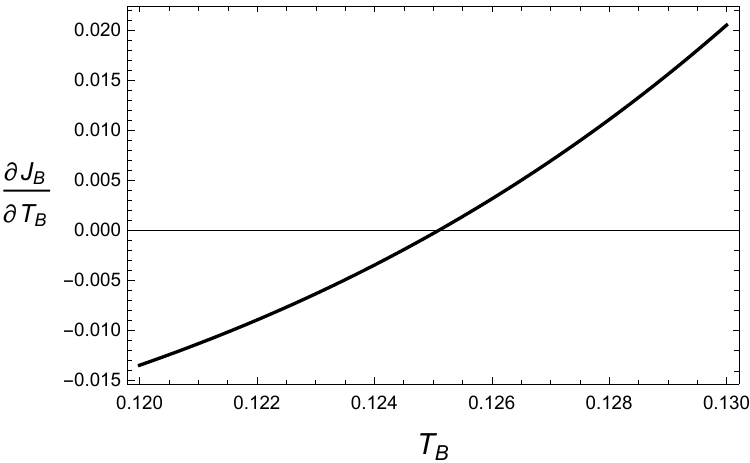}
\caption{Steady-state transistor-effect amplification factor can diverge. To demonstrate this, we plot here, against \(T_B\), the quantity $\frac{\partial J_{B}}{\partial T_{B}}$, which appears in the denominator of the amplification factor, \(\alpha\), for the steady state quantum thermal transistor. We choose \(T_A = 0.2\), and \(T_C=0.02\), as in Ref.~\cite{Tran4}. The horizontal axis is dimensionless, while the vertical one is in units of \(\hbar\delta^2\).}
\label{fig:der}
\end{figure}

 The heat current is defined as the amount of entropy exchanged per unit time between the open system and the environment. The flow of heat current is caused by the change of the internal energy $E =\text{Tr}(H_{sys}\rho)$, which results from dissipative effects \cite{Petruccione2002}. So, heat current can be expressed as 
\begin{equation}
J_{X}=\text{Tr}(H_{sys} \mathcal{L}_{X}[\rho]).
%&=& -\frac{\hbar \delta}{K_B T_{X}} \frac{K_B}{2} \text{Tr}(\sigma_{z}^{X} \sigma_{z}^{Y} \mathcal{L}_{X}[\rho])
\label{HC}
\end{equation}
%\textcolor{blue}{
Let us define $T_{X}=\frac{K_B \tilde{T}_{X}}{\hbar \delta}$.
%}
Now, we briefly discuss the salient points, for our purposes, in the operation of a QTT, along the description in \cite{Tran4}. For lower values of $T_{B}$, a small change in $J_{B}$ results in a relatively large change in $J_{A}$ and $J_{C}$, giving rise to the transistor effect. This is analogous to a semiconductor transistor where a change in the base current controls the flow of emitter and collector currents. In the case of the QTT, the dynamic amplification factor $(\alpha)$ is defined as 
\begin{center}
\begin{equation}
\alpha_{X}=\frac{\partial J_{X}}{\partial J_{B}}=\frac{\frac{\partial J_{X}}{\partial T_{B}}}{\frac{\partial J_{B}}{\partial T_{B}}}
\label{eq:amplification}
\end{equation}
\end{center}
where $ X=A,C$.
%In the literature, the dynamic amplification is defined by the change of $J'_{A,C}=-\tilde{T}_{A,C}J_{A,C}$ with respect to a change in $J'_{B}=-\tilde{T}_{B}J_{B}$\cite{Tran4,a22}. To keep it consistent with the units and the definition given in Eq. (\ref{HC}), we have used the modified definition of the dynamic amplification.  
It provides a measure of the efficiency of the system in working as a transistor.
For the quantum thermal transistor to work, the heat flow between one of the three TLSs and the bath that it is coupled with (in our case, \(J_B\)) should control the other two heat currents (\(J_A\) and \(J_C\)). This would require that a small change in \(J_B\) should result in  relatively large changes in \(J_A\) and \(J_C\), so that we have  \(|\alpha|>1\). A higher value of \(|\alpha|\) would mean that significantly better control of \(J_A\) and \(J_C\) has been achieved. Similar measures for efficiency have also been used in Refs.~\cite{Tran4,Tran5,ex1,Tran8,Tran7,Tran6,ex2,ex3,ex4,n2,ex6}.
This amplification factor is analogous to the base transport factor $(\beta=\frac{\Delta I_C}{\Delta I_B})$, which is the ratio of change in collector to change in base current in the classical transistor. For a good transistor, $\beta \approx 100$, whereas in the quantum case, the three TLS-bath setup behaves like a transistor if $|\alpha_X|>1$ and a good one if $|\alpha_X|$ has a sufficiently large value. Note however that the $\beta$ for a classical transistor and $\alpha$ for its quantum counterpart are not directly comparable. The elements of $\alpha$ are the heat currents corresponding to the exchanges of heat between 
%the bath and 
the TLSs and their respective baths, and is not a result of the movement of any particle (electrons in case of a classical transistor). For the steady state situation, $\frac{d\rho}{dt}=0$, thereby yielding $\sum_{X=A,B,C} J_{X}=0$. So, $\alpha_A+\alpha_C+1=0$ and when $\alpha_A$ and $\alpha_C$ have large values, then $\alpha_{A} \approx -\alpha_{C}$.  This feature is lacking in the case of general evolved states.
Furthermore, for certain values of $T_{B}$, the dynamic amplification attains very high values \cite{Tran4}. For a steady state quantum thermal transistor, $\alpha$ can reach to infinity for a certain value of $T_B$. This is due to the fact that, at such points, the denominator, $\frac{\partial J_{B}}{\partial T_{B}}$, of the expression for \(\alpha\) vanishes (with the numerator remaining finite). This is reminiscent of the Coulomb blockade where the electrical conductance vanishes at low bias voltages.
%which is the denominator of the expression of $\alpha$, at such points. 
See Fig. \ref{fig:der}.

In this paper, we have obtained the partial derivatives of the numerator and denominator of Eq. (\ref{eq:amplification}) numerically, using the \textit{five-point (midpoint) numerical formula}. The first-order derivative of a function $f(x)$ at point $x_0$ is then given as
\begin{eqnarray}
%\resizebox{.42 \textwidth}{!} 
%{
f^{\prime}(x_0)=\phantom{rupasi bijalilata pareni ta mante ami tomar aa}\nonumber\\\frac{f(x_0-2h)-8f(x_0-h)+8f(x_0+h)-f(x_0+2h)}{12h}.\phantom{ab}
%}
\end{eqnarray}
An error of order $h^4$ appears here, which we neglect for \(h\) of the order \(0.001\). 
%Since we have fixed all the parameters other than $T_{B}$, this can be used for calculation the partial derivatives as well. 
In our case, the numerator and the denominator of Eq. (\ref{eq:amplification}) are functions of $T_B$, and $T_A=0.2$ and $T_C=0.02$ as in \cite{Tran4}.
\section{A quantum transistor in transient regime}
\label{sec:transient}
A steady state QTT functions efficiently in the region $0<T_{B}<0.15$, where a slight change in $J_{B}$ results in large changes in $J_{A}$ and $J_{C}$; this gives us high values of $\alpha$. 
%The denominator $\frac{\partial J_{B}}{\partial T_{B}}$ in Eq. (\ref{eq:amplification}) reduces and reaches  zero at $T_B\approx 0.12$ and then increases further. 
$\frac{\partial J_{B}}{\partial T_{B}}$ reduces with respect to \(T_{B}\), for which the amplification factors shoot to infinity at \(T_{B}\approx 0.12\).
Therefore, at $T_B\approx 0.12$, the magnitude of amplification factor shoots to infinity and then again reduces; it falls to nearly equal to zero after $T_{B}>1.15$ and the system no longer works as a transistor. See \cite{Tran4} for a detailed discussion of the quantum thermal transistor operating in the steady state regime.\par
It is possible that the time taken by the system to reach the steady state is quite large and continuing the time evolution for such a long duration may result in some noise or fluctuations in the tuning parameters of the setup. This may result in straying 
%So, we might go 
outside the ideal working parameters and the system may no longer serve our purpose. This could be overcome by operating the QTT in the transient regime, where the time of evolution of the system is smaller than the time taken to evolve into the steady state and the operation of transistor is better or at least as good as the steady state regime. Such a QTT could prove to be advantageous over the same operating in the steady state regime. But it 
%this 
is not 
%necessarily 
true that, from every initial state, such a transient QTT could be obtained.\par 
In this paper we discuss some cases where the transient regime is beneficial over the steady state of a QTT. We have taken different initial states and studied their amplification properties in the transient regime. We find that there exist several states that follow a nature similar to that of the steady state QTT. For these initial states, transistor effect is visible even for smaller times. There are also some states for which the dynamic amplification dies out for much smaller values of $T_{B}$. These states are not very good for our purposes, but not useless either, because, although the operating region of the transistor is $T_B<0.15$, the amplification factors have much larger values; these are better transistors for a very small region of $T_B$. Nonetheless, there exist states for which the transistor effect is retained even after $T_{B}\geq 0.2$. These are of particular interest as they promise more efficiency than that of the steady state QTT without the long wait to reach the steady state. In the next subsections, we discuss some such initial states and the corresponding transistor effect they show.

\subsection{Paradigmatic initial states for reaching the transient regime}
\label{sec:ghz}
\begin{figure*}
\includegraphics[width=\textwidth]{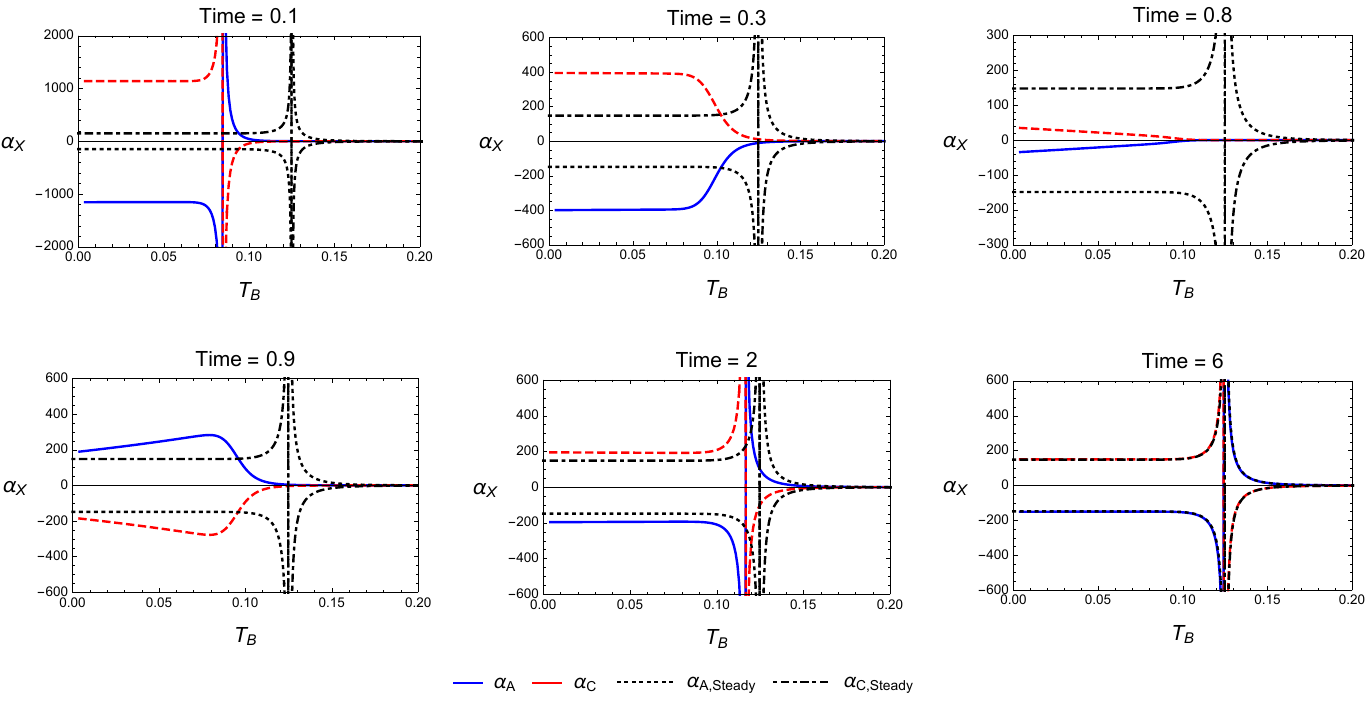}
\caption{Transient three-qubit quantum thermal transistor with the time evolution beginning in the Greenberger-Horne-Zeilinger (GHZ) state. The panels exhibit the nature of amplification factors with the change of $T_B$ for different dimensionless times. The initial state is taken as the $GHZ$ state. The axes represent dimensionless parameters in all panels. We remember that time is used in this paper in the unit of \(\delta\), so that the times mentioned in the panels when divided by \(\delta\) has the unit of time.}
\label{fig:ghz}
\end{figure*}
We start with the well-known three-qubit Greenberger-Horne-Zeilinger state ($GHZ$ state \cite{ekhane-GHZ}), which has the form
\begin{equation}
 \ket{GHZ}=\frac{\ket{000}+\ket{111}}{\sqrt{2}}.
\end{equation}
The nature of the amplification factor for different times with the $GHZ$ state as the initial state in the time evolution is illustrated in Fig. \ref{fig:ghz}. As is evident from the plots, for times satisfying $0.1 \leq t \leq 0.3$, the amplification factor is more than that of the steady region QTT, but only for $T_{B}<0.8$. 
We remember that time is used in this paper in the unit of \(\delta\), so that \(t/\delta\) has the unit of time.
If we further increase time, the amplification factor steadily decreases to a very small value; at $t \approx 0.8$, the transistor is not working at all, after which the amplification factor increases again; it decreases once more and coincides with the steady state when $t \approx 6$, after which it remains the same. Thus we can have a good transient QTT with better efficiency than the steady state QTT but for a reduced range of $T_{B}$ .

\par Another paradigmatic genuine three-party entangled 
%non-biseparable 
three-qubit 
%entangled 
state is the $W$ state \cite{ekhane-W1, W} (see also \cite{ekhane-W2}), 
%which has the following form
\begin{equation}
\ket{W}=\frac{\ket{001}+\ket{010}+\ket{100}}{\sqrt{3}}.   
\end{equation}
Since the evolution of the $W$ state follows a trend similar to that of the $GHZ$ state, the rest of the discussion about the \(W\) state is included in Appendix \ref{W,000}. A discussion about 
%along with the discussion on 
the state $\ket{000}$ as the initial state is also given in the same appendix.
\par
\begin{figure*}
\includegraphics[height=6cm,width=15cm]{
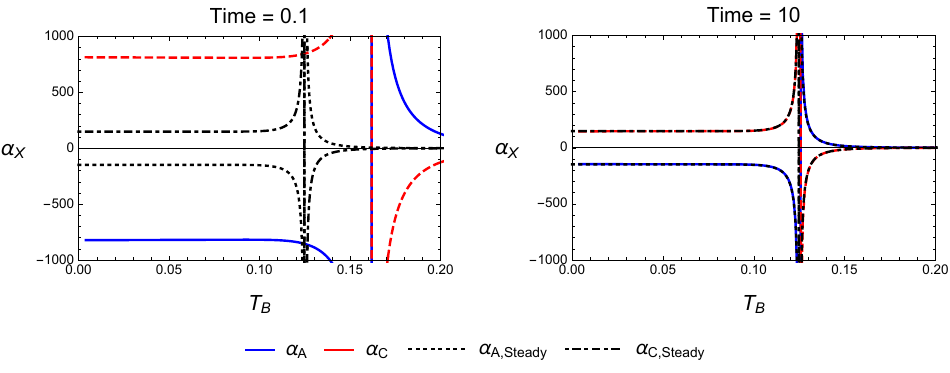}
\caption{Nature of amplification factors with the change of $T_B$ for different dimensionless times. The initial state is taken as $\ket{001}$. Both  axes represent dimensionless parameters in all panels. For the same cases at $Time=3$ and $Time=6$, see Appendix \ref{extra}.}
\label{fig:001}
\end{figure*}

\begin{figure*}
\includegraphics[height=6cm,width=15cm]{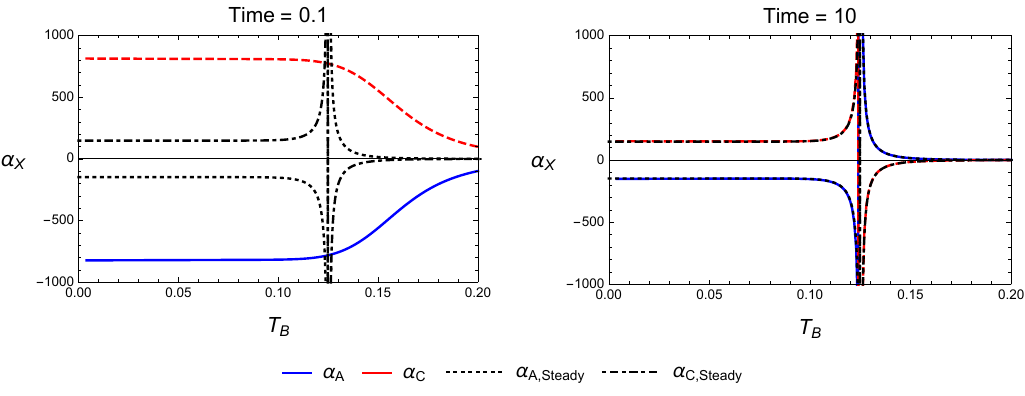}
\caption{Nature of amplification factors with the change of $T_B$ for different dimensionless times. The initial state is taken as $\ket{011}$. Both  axes represent dimensionless parameters in all panels. For the same cases at $Time=3$ and $Time=6$, see Appendix \ref{extra}.}
\label{fig:011}
\end{figure*}

 The other two initial product states we have studied are $\ket{001}$ and $\ket{011}$. For these, we get appreciably good QTT which has a significant advantage over a steady state QTT beyond $T_B \geq 0.2$. This advantage in the behavior of the QTT when $\ket{001}$ is chosen as the initial state is shown in Fig. \ref{fig:001}. For $t=0.1$, we have a very large amplification factor as well as a larger operating region of $T_B$. As time increases, the amplification factor reduces and ultimately reaches the steady state for times given by $t \approx 10$. (Please refer to Appendix \ref{extra} for the plots corresponding to certain  intermediate times, viz. $t = 3$ and $t = 6$.) During the time evolution, it remains better than the steady state QTT, and it always gives a larger operating region of $T_B$ until it goes to the steady state regime. For 
%the initial state 
$\ket{011}$ as the initial state (see Fig. \ref{fig:011}), the transient transistor gives better efficiency and a larger operating region of $T_B$ than that of the steady state transistor for small times ($t=0.1$). For further increase of time, the amplification factor reduces. At $t=3$, it provides a slightly larger operating region with a lesser amplification factor than the steady state transistor. For $t=6$ (see Appendix \ref{extra}), it is not at all a better transistor than the steady state one and, for $t \approx 10$, it reaches to the steady state case.  For both the  cases, the transient transistors can be considered as necessarily transient QTT for $t=0.1$. Some more examples of necessarily transient transistors are discussed in Sec. \ref{sec:ntrans}.
\subsection{Random initial states}
\label{sec:random}
\begin{figure*}
\includegraphics[width=\textwidth]{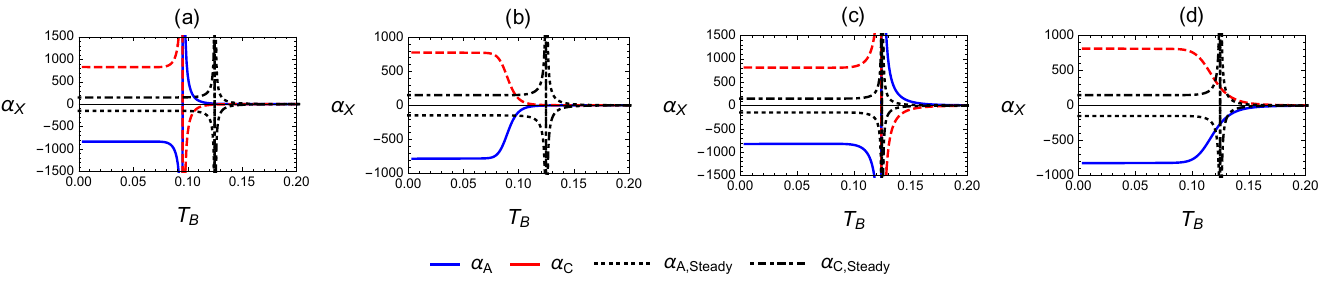}
\caption{Nature of amplification factors with the change of $T_B$ for different randomly chosen states. See text for details. Both axes of all panels represent dimensionless parameters.}
\label{fig:random}
\end{figure*}
Until now, we have studied the QTT in the transient regime by considering a few paradigmatic three-qubit states as initial states of the time evolution. 
%It is not enough to study only a few initial states for finding a good transient transistor. There are a large number of states whose nature in the transient regime can be explored. 
There are of course a large number of other states that could be explored. It is not clear now if we should classify them for their usefulness in the QTT task. With respect to their entanglement content, these states can 
%These can be 
broadly be divided into product, biseparable,  $GHZ$- and $W$-class  states \cite{W}. We expect that exploring random elements of these classes of states would provide an intuitive picture of approximately the entire space of states. We analyzed 350 Haar uniformly generated states belonging to either one of these classes.\par
The three-qubit genuinely multiparty entangled pure states are all $GHZ$-class states except for a set of measure zero \cite{W}. 
%some special cases. 
The general form of the $GHZ$-class state is therefore the same as the general three-qubit pure state, up to a set of measure zero, and the latter is given by the following, which we christen as \(|\psi\rangle_{GHZ}\):
\begin{eqnarray}
\nonumber
\ket{\psi}_{GHZ}=a\ket{000}+b\ket{001}+c\ket{010}+d\ket{100}\\
+a_1\ket{011}+b_1\ket{110}+c_1\ket{101}+d_1\ket{111},
\label{eq:ghz}
\end{eqnarray}
where $a,\,b,\,c,\,d,\,a_1,\,b_1,\,c_1,\,d_1$ are the  complex coefficients, constrained by the normalization condition. 
%i.e., 16 real numbers. So, to randomly generate a $GHZ$ class state we have to generate 16 independent random numbers from Gaussian distribution with mean $0$ and standard deviation $1$ simultaneously. By this method we can generate $GHZ$ class states Haar uniformly.
\par
For generating $W$-class states Haar uniformly, we take a different  approach. The general form of a $W$-class state is \cite{W} 
\begin{equation}
\ket{\psi}_{W}=a\ket{001}+b\ket{010}+c\ket{100}+d\ket{000},
\label{eq:W}
\end{equation}
and states that are local unitarily connected to it, where $a,\,b,\,c,\,d$ are arbitrary real numbers up to the normalization constant. 
%See \cite{W} for a discussion on generating $GHZ$ and $W$ state. The $W$ class states form a set of measure zero in the set of all pure three-qubit states. To generate the $W$-class states we have to Haar-uniformly generate the above state with real $a,\,b,\,c,\,d$ and then independently and Haar-uniformly generate the three local unitaries. 
As an easier alternative to the Haar-uniform generation of the \(W\)-class states, we proceed as follows:
%this would be the following:
%\begin{itemize}
(a) generate the above state with \textit{complex} $a,\,b,\,c,\,d$;
(b) generate the above state with complex $a,\,b,\,c,\,d$ but with $\ket{0}$ and $\ket{1}$ replaced by $\sigma_x$ eigenvectors;
(c) repeat the same using $\sigma_y$ eigenvectors.
%\end{itemize}
%In each case we have to generate 8 independent real numbers from Gaussian distribution with mean $0$ and standard deviation $1$ simultaneously.
In this way, we have a fair chance of having a picture of the entire \(W\) class.\par  
In the biseparable class of states, two out of the three TLS are entangled in each state. There can therefore be three types of biseparable states: separable in the $A:BC$ cut, or the $B:AC$ cut,  or the $AB:C$ cut. The general form of a biseparable state that is separable in the $AB:C$ cut is
\begin{equation}
\ket{\psi}_{AB:C}=(a\ket{00}+b\ket{01}+c\ket{10}+d\ket{11}) \otimes (a_1\ket{0}+b_1\ket{1}). 
\label{eq:12:3}
\end{equation}
Similarly for the $B:AC$  and $A:BC$ cuts. 
%So, for this case we have to generate $6$ complex numbers i.e., $12$ real numbers randomly as in the previous cases.
\par
The next (and final) is the class of $3$-qubit product states which have the general form
\begin{equation}
\ket{\psi}_{Product}=(a\ket{0}+b\ket{1})\otimes (a_1\ket{0}+b_1\ket{1}) \otimes (a_2\ket{0}+b_2\ket{1}).
\label{eq:Product}
\end{equation}
%As in case of the biseparable states, the product state also requires us to generate $12$ real numbers randomly. 
We studied $50$ Haar-uniformly generated states each from the classes $GHZ$, the three types of $W$-class states mentioned above, biseparable states separable in the $A:BC$ cut, biseparable states separable in the $AB:C$ cut, and product states, making a total of $350$ states. We set the time so that $t=0.1$, because many of these states reach their steady states rather quickly. Moreover, having a transient QTT at a short time is advantageous.\par
For these $350$ initial states, we get three types of transient QTTs. One of them 
%type of transient transistor 
has the same operating efficiency as discussed in Sec. \ref{sec:transient} for $GHZ$, $W$ and $\ket{000}$ as initial states. They have a good amplification property in a smaller region of $T_B$, but have a better efficiency than that of the steady state transistor. This type of transient QTT is depicted in panels (a) and (b) of  Fig. \ref{fig:random}.
%$(a)$ and $(b)$. 
The second type of transient transistor is shown in the last two panels of Fig. \ref{fig:random}.
%$(c)$ and $(d)$. 
These have an operating region of $T_B$ that is similar to that of the steady state transistor, but has a larger value of amplification factor, which makes us conclude that these types of transistors are more advantageous than a steady state transistor. The third type of transient transistor is one of the main interests of our paper, and is discussed in the succeeding section. Note that we can categorize the transient transistors depending on their operating region with respect to $T_B$, but we cannot categorize them depending on the chosen class from which the initial state is used. The three types of transistors can be found in any one of the classes of states we have mentioned above. This indicates that the initial entanglement content of the system seems to have little to do with its efficiency for the quantum task chosen at hand.
\section{Necessarily transient quantum transistor}
\label{sec:ntrans}
%In this section, we discuss our findings on  The most interesting results from our so long observations are discussed in this section. 
We found that, for certain states, 
%a transistor functions better in the transient region before evolving into the steady state. For such states, 
we get significant amplification even for $T_{B}\geq 0.2$, where the transistor no longer functions in the steady state regime. As a consequence, these types of transistors can be considered as \textit{necessarily transient quantum transistors}, and potentially form a beneficial quantum device not available in the steady state regime. Such states can belong to either of the four classes of states discussed in the preceding subsection, viz. the \(GHZ\) and \(W\) classes, biseparable states and product states. 
The transistors using
%In addition to these the 
initial product states $\ket{001}$ and $\ket{011}$ can also be included in this category of necessarily transient ones, as already discussed in Sec. \ref{sec:transient}. Now, we shall discuss four such initial states and the operating regions of their providing transient thermal transistors. It may be noted here that within the literature of quantum refrigerators, transient refrigerators were considered in Refs. \cite{R5, R7} and necessarily transient refrigerators were reported in Ref. \cite{Tr1}.\par
\begin{figure*}
\includegraphics[width=\textwidth]{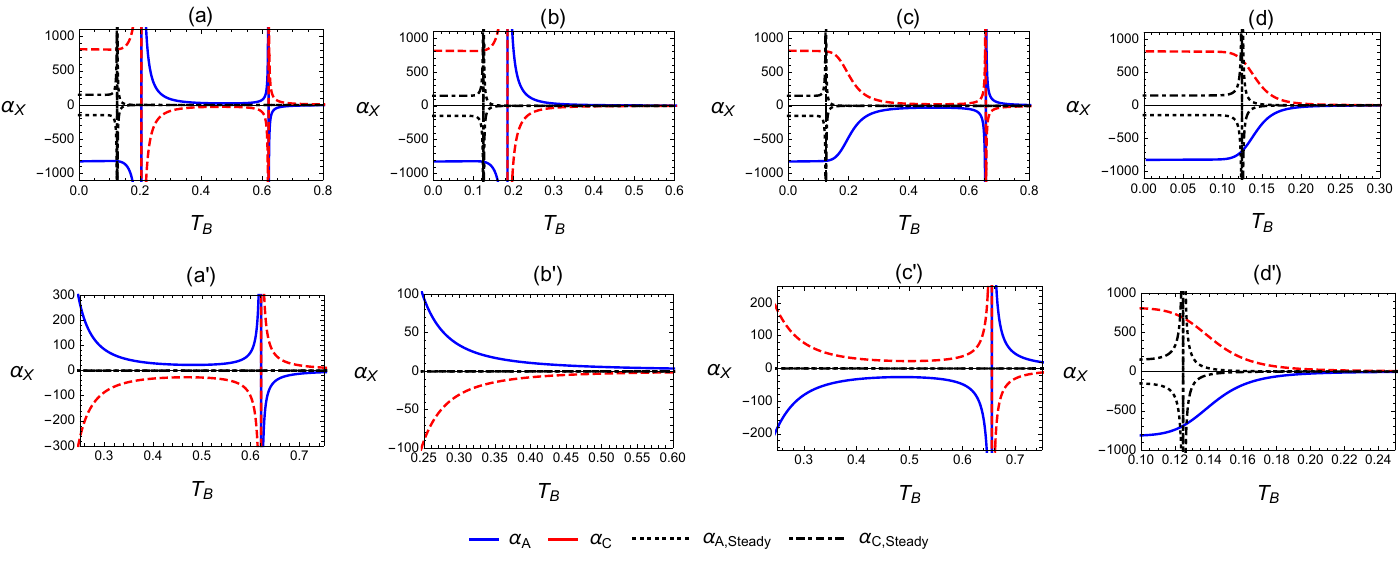}
\caption{Necessarily transient quantum transistors. Operating regions of necessarily transient transistors are exhibited for the initial states chosen from four classes of states: $(a)\;\;\ket{\psi^{\prime}_{GHZ}}$, $(b) \;\;\ket{\psi^{\prime}_{AB:C}}$, $(c) \;\;\ket{\psi^{\prime}_{W}}$, and $(d)\;\;\ket{\psi^{\prime}_{Product}}$. To visualize the operating region beyond the capability of a steady state transistor, magnified views are shown in $(a^{\prime}),\,(b^{\prime}),\,(c^{\prime}),\,(d^{\prime})$, respectively, in the relevant regions. All axes represent dimensionless parameters.}
\label{fig:nt}
\end{figure*}
One such initial state is taken from the Haar-uniformly generated $GHZ$-class 
%of 
%random 
states, denoted as $\ket{\psi^{\prime}_{GHZ}}$, and which has the form  in Eq. (\ref{eq:ghz}).
The amplification factor for $0<T_B\leq 0.8$ is shown in Fig. \ref{fig:nt} $(a)$. It is clearly visible that in this case, the transient QTT has much better amplification in the region $0<T_B<0.15$ than that of a steady state transistor. Moreover, the $GHZ$-class transient transistor has also a good amplification beyond $T_B=0.15$. The efficiency of the transistor in $0.25 \leq T_B \leq 0.75$ is magnified in Fig. \ref{fig:nt} $(a^{\prime})$. The amplification is not very good in $0.4 \leq T_B \leq 0.6$, but 
it 
%the amplification 
is still about $10$. This is not appreciably good, but if we are constrained to be in the region $0.3 \leq T_B \leq 0.6$, then a transient QTT is possible using this state, although a steady-state one cannot be used.\par
The next example of a necessarily transient quantum transistor is obtained from an initial biseparable state $\ket{\psi^{\prime}_{AB:C}}$, that is  separable in the $AB:C$ cut, and has the form in Eq. (\ref{eq:12:3}).
This case is depicted in Fig. \ref{fig:nt} $(b)$, and this shows a nature similar to the previously-discussed $GHZ$-class transient transistor. The major difference is that in the $GHZ$ case, after reducing near $T_B=0.3$, the amplification factor again rises and reaches an appreciably large value for $T_B > 0.6$, but in this biseparable case, the amplification factor reduces near $T_B=0.4$ and does not increase again. This is clear in Fig. \ref{fig:nt} $(b^{\prime})$, which is a magnified version of Fig. \ref{fig:nt} $(b)$. The operating region of this transient transistor is $0 < T_B < 0.4$, which is not bad at all.\par
Necessarily transient quantum thermal transistor can also be obtained from the $W$-class of states. One example of such a state is $\ket{\psi^{\prime}_{W}}$, having the form in Eq. (\ref{eq:W}), with the corresponding results being given in Fig. \ref{fig:nt} $(c)$.
From Fig. \ref{fig:nt} $(c)$ and its magnified version $(c^{\prime})$, we can see that the operating region of the transient transistor having the initial state $\ket{\psi^{\prime}_{W}}$ is $0<T_B<0.7$ which is the same as that of the necessarily transient transistor having the initial state $\ket{\psi^{\prime}_{GHZ}}$ that we have already discussed. The long-$T_B$ behavior is qualitatively the same in these two cases (compare Figs. \ref{fig:nt} $(a^{\prime})$ and $(c^{\prime})$) but the short-$T_B$ behavior is interestingly different (compare Figs. \ref{fig:nt} $(a)$ and $(c)$). Despite this qualitative difference at short \(T_B\), the quantity of amplification provided by necessarily transient transistors having $\ket{\psi^{\prime}_{GHZ}}$ and $\ket{\psi^{\prime}_{W}}$  as their initial states are almost the same before they shoot to %\textcolor{blue}{
a very large value
%} 
and go to zero respectively.\par  
The last one we discuss here is a necessarily transient transistor having a product state as its initial state. The product state $\ket{\psi^{\prime}_{Product}}$ has the structure as in Eq. (\ref{eq:Product}). 
Figures \ref{fig:nt}$(d)$ and \ref{fig:nt}$(d^{\prime})$ [magnified version of $(d)$] show that this necessarily transient transistor provides a similar nature of amplification factor to that of the $W$-class state for small $T_B$, but as in $\ket{\psi^{\prime}_{AB:C}}$ discussed above, the rising of the amplification factor twice is missing here. This transient transistor has the smallest operating region $0< T_B < 0.18$ compared to the preceding three, but can be categorized in the necessarily transient transistor. 
\par 
The exact forms of the four initial states %which we have taken as 
which we have considered for the analyses in the figures, viz. 
$\ket{\psi^{\prime}_{GHZ}}$, $\ket{\psi^{\prime}_{AB:C}}$,  $\ket{\psi^{\prime}_{W}}$, $\ket{\psi^{\prime}_{product}}$, are given in Appendix \ref{values}. For these four initial states, we can structure a necessarily transient transistor. 
\par
A note about the preparation of the initial states is in order here. The product state, $\ket{\psi^{\prime}_{product}}$, can be created by locally magnetizing the three spins-1/2 system in the appropriate directions. This requires  local preparation of the states of each two-level system separately, and can be done on almost all physical platforms. The other three states can be prepared by local filtering operations on the \(GHZ\), \(W\), or two-qubit maximally entangled states (``Bell'' states). %Creating such types of initial states from the entire family of the four classes of states is quite tricky. One can try local filtering operations for creating such type of initial states. 
See \cite{filter1, filter2} in this regard.
Preparation of \(GHZ\), \(W\), and Bell states has been attained in several physical systems,
and requires non-trivial entangling operations, whose exact form depends on the physical system used.
See, e.g., Refs.\cite{add1,add2,add3,add4,add5,add6,add7,add8,add9}, and references therein. 
\par
There are many states with this advantageous ability, but we discussed only four of them. It is important to note that all the necessarily transient QTTs have the amplification properties that are similar to any of these four. In our examples, we found that the transient transistors having the initial states $\ket{\psi^{\prime}_{GHZ}}$ and $\ket{\psi^{\prime}_{W}}$ illustrated in Figs. \ref{fig:nt}$(a)$ and \ref{fig:nt}$(c)$ as the most advantageous compared to other examples we have discussed. 
This does not mean that all necessarily transient transistor having $GHZ$- and $W$-class states as their initial states always have the better advantage over the other two classes. This fact is clearly visible from Figs. \ref{fig:001} and \ref{fig:011}. Both are necessarily transient transistors having two different initial product states, but the behavior of the amplification factors are completely different, and yet the operating region is almost the same. The difference in operating regions of  necessarily transient transistors having initial product states can be understood by comparing Figs. \ref{fig:001}, \ref{fig:011}, and \ref{fig:nt}($d$). We can conclude that the four panels in Fig. \ref{fig:nt} are not the representatives of their corresponding classes, but they are the representatives of the four types of necessarily transient quantum thermal transistors.
\subsection{Variation of amplification factor with time}
\label{sec:time_vary}
\begin{figure*}
\includegraphics[height=10cm,width=12cm]{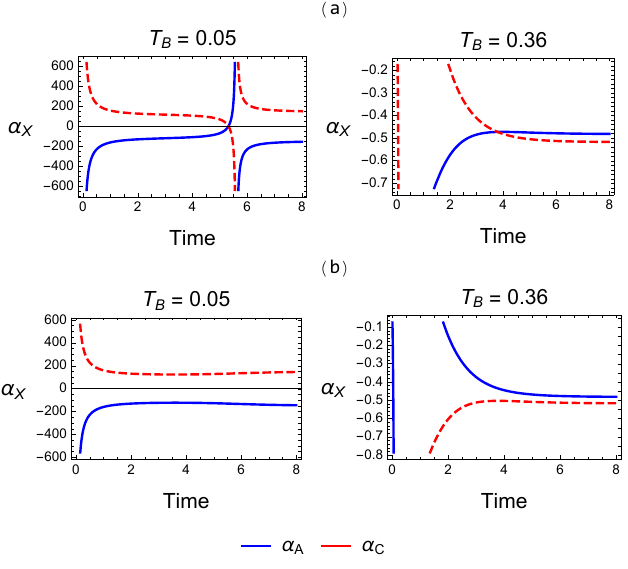}
\caption{Time evolution of amplification factor for different $T_B$ for initial states taken as $(a) \;\;\ket{\psi^{\prime}_{GHZ}}$ and $(b)\;\;\ket{\psi^{\prime}_{W}}$. All the axes represent dimensionless parameters. In Appendix~\ref{extra}, we present plots also for $T_B = 0.13$ and $T_B = 0.26$.}
\label{fig:time}
\end{figure*}
Until now in this section, 
%In the preceding section, 
we discussed some cases of necessarily transient QTTs at $t=0.1$. At this time, the system studied remains in transient regime and is far  from its steady state sector. In all previous plots of amplification factors at different times, we observed that all the transistors are in their transient regime at $t=0.1$, and we get better efficiency around that time compared to later times. But it is necessary to identify the 
range of 
%for which values of 
time for which we can have a better transistor than the steady state one. Thus observing the time evolution of this system is also crucial before we can consider it as a transient QTT. It also helps us to have an idea about how fast the system reaches the steady state regime. We have already discussed for which value of time the transient transistors having initial states as $GHZ$, $W$ and $\ket{000}$ provide good efficiencies, and that the necessarily transient transistors having initial states $\ket{001}$ and $\ket{011}$ give better efficiencies as well as a sufficiently large operating region in $T_B$, in Sec. \ref{sec:transient}. Here we analyze the time evolution of the amplification factor of the necessarily transient transistor having the initial states $\ket{\psi^{\prime}_{GHZ}}$ in Fig. \ref{fig:time} $(a)$, and  having $\ket{\psi^{\prime}_{W}}$ as the initial state in Fig. \ref{fig:time} $(b)$. We have studied the time evolution by fixing $T_B$ at four values, viz. $0.05,\,0.13,\,0.26$ and $0.36$. (Plots for $T_B = 0.13$ and $T_B = 0.26$ are in Appendix~\ref{extra}.) The values of $\alpha_A$ and $\alpha_C$ for these values of $T_B$ for a steady state transistor are given below:
\begin{eqnarray}
T_B&=&0.05: \quad \alpha_A\approx -149, \quad \phantom{.}\alpha_C \approx \phantom{+}148; \nonumber\\
T_B&=&0.13: \quad \alpha_A\approx \phantom{+}206, \quad \phantom{.}\alpha_C \approx -207; \nonumber\\
T_B&=&0.26: \quad \alpha_A\approx-0.4, \quad \phantom{2}\alpha_C \approx -0.6; \nonumber\\
T_B&=&0.36: \quad \alpha_A\approx-0.5, \quad \phantom{2}\alpha_C \approx -0.5. \nonumber
\end{eqnarray} 
We are interested in those times for which we can get a better or at least the same efficiency as in the steady state case. At $T_B=0.05$, the transistor having the initial state $\ket{\psi^{\prime}_{GHZ}}$ has a good efficiency (better or the same as the steady state efficiency) for $0 < t \leq 3$, very close to $t=6$, and for the further times where it has already reached the steady state (for our purposes). So, it is good to stay in the region $0 < t <1$ and very close to $t=6$ for achieving a good amplification. For 
%the 
%corresponding 
$\ket{\psi^{\prime}_{W}}$ as the initial it is better to stay in the region $0 < t <1$. At $T_B=0.13$, for both the initial states, $\ket{\psi^{\prime}_{GHZ}}$ and $\ket{\psi^{\prime}_{W}}$, it is  better to stay in $0 < t <1$. For $\ket{\psi^{\prime}_{W}}$, it is also good to stay at $t$ close to $6$. For $t > 6$, for both the states, the transistor reaches the steady state operating regime. At $T_B=0.26$ and $t=0.36$, we can see that for both the cases, it is better to stay at times near zero, because for a very small time, the amplification factor reduces very fast and goes to a very small magnitude ($0.5$). Hence, for the two most useful necessarily transient QTTs, we see that we can get better efficiencies if we stay at times  close to zero for all $T_B$. In all our previous studies (in this section), we were at $t=0.1$, which is closer to zero, and so providing a good amplification factor.
\section{Relation between $\alpha_A$ and $\alpha_C$}
\label{sec:diff_alpha}
\begin{figure*}
\includegraphics[width=\linewidth]{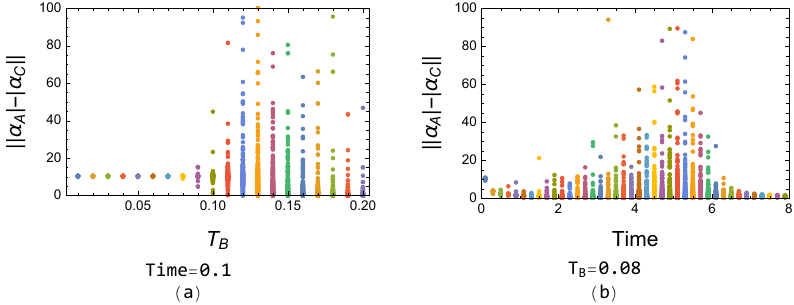}
\caption{Variation of $||\alpha_A|-|\alpha_C||$ with $(a)\;\;T_B$ for time fixed at $t=0.1$, and \((b)\;\;\) time for $T_B$ fixed at $0.08$. All the axes represent dimensionless parameters.}
\label{fig:scatter}
\end{figure*}
We have already stated the relation between $\alpha_A$ and $\alpha_C$ for  transistors operating in the steady state regime in Sec. \ref{sec:tools}. The relation, $\alpha_A+\alpha_C+1=0$, is however not valid in the transient regime. For transient sectors, the relation between $\alpha_A$ and $\alpha_C$, obtained from Eq. (\ref{eq:dynamics}), is given by
\begin{equation}
\alpha_A+\alpha_C+1=\Big(\frac{\partial J_{B}}{\partial T_{B}}\Big)^{-1}\text{Tr}\Big[H_{sys}\frac{\partial}{\partial T_{B}}\Big(\frac{d \rho}{dt}\Big)\Big{]}. 
\label{alpha_tran}
\end{equation}
For the steady state regime, $\frac{d \rho}{dt}=0$, whereby the right-hand-side (RHS) of Eq. (\ref{alpha_tran}) reduces to zero. Till now, we were concentrating on both the amplification factors together and found that when amplification is low then both $\alpha_A$ and $\alpha_{C}$ are low, and when amplification is high then both are high, but we did not have any idea whether $\alpha_A$ and $\alpha_{C}$ have almost the same magnitudes in sectors where they are very high. For the steady state case, it is always true that $\alpha_A \approx \alpha_C$ when amplification is high, but what is the scenario for the transient regime? This is the question that we analyze in this section. The difference of the magnitudes of $\alpha_A$ and $\alpha_{C}$ for different values of $T_B$ is depicted in Fig. \ref{fig:scatter}$(a)$, and the same for different values of time is depicted in Fig. \ref{fig:scatter}$(b)$. These are the scattered plots of $||\alpha_A|-|\alpha_C||$ for the $350$ initial states chosen randomly. From Fig. \ref{fig:scatter}$(a)$,  we can infer that, when $T_B$ is low, the differences of the magnitudes of  $\alpha_A$ and $\alpha_{C}$ are almost the same for all states and it is not very high (being $\approx 10$). In all the  cases studied, we have seen that, for low $T_B$ and at $t=0.1$, amplification factors are sufficiently high and so the difference in magnitudes of the amplification factors was not visible in the previous plots. Now, for further increase of $T_B$ the probability of having a greater difference in magnitudes of the amplification factors increases and for $0.12 \leq T_B \leq 0.15$ there is a finite probability to have this difference very high. For $T_B=0.15$, we have got the maximum $||\alpha_A|-|\alpha_C|| \approx 6000$ with a very small probability. 
%but always t
The probability of having a smaller difference in magnitudes is greater. We have set the range of $||\alpha_A|-|\alpha_C||$ axis from $0-100$ to show that the probability of having a smaller difference is always greater but there is a finite probability of having a larger difference for $T_B$ near $0.12-0.15$, which is not shown in this figure. This large difference is also not visible in the previous plots because for this region of $T_B$, amplification factors are very large and the range of the $\alpha_X$ axes were not so large. If we further increase $T_B$, the probability of having greater $||\alpha_A|-|\alpha_C||$ reduces and, for 
%maximum 
most
states, it comes to below $20$.\par
Next we discuss how the difference $||\alpha_A|-|\alpha_C||$ changes with time for a fixed $T_B$. We have fixed \(T_B\) at \(0.08\), because from the preceding observations, we have found that at $T_B=0.08$, all the transient thermal transistors have a good amplification. From Fig. \ref{fig:scatter} $(b)$, we can see that the differences of the magnitudes of the two amplification factors for different times at a fixed $T_B$ follows a similar nature as they have with different $T_B$ at fixed time. For smaller times, all the $350$ states have very small differences between $\alpha_A$ and $\alpha_{C}$ in magnitude. So, in the cases where we were at $t=0.1$, 
%for all the above discussed cases, so 
we got $\alpha_A \approx \alpha_C$ for $T_B=0.08$. Now, if we go ahead on the time axis, the probability of having greater $||\alpha_A|-|\alpha_C||$ increases and, for $2 < t \leq 6$, we can have a finite probability to have some initial states for which this difference is very high. The maximum $||\alpha_A|-|\alpha_C||$ we have obtained is $\approx 60000$ for $t=4.8$. For $t > 6$, the differences reduce for all states and are very near to zero, implying – within Haar-uniform generation accuracy and  within the set of physical quantities studied – that all states reach the steady state regime for $t > 6$.   

\section{Physical realisation}
\label{phys}
In this short section, we aim to provide indications to potential physical systems where the phenomenon discussed can be implemented, and also discuss the corresponding numbers in real units. 
%It is also important to give a physical realisation of a theoretical model. 
There are some previous works in which proposals of experimental setups for similar ends are given. See Refs.~\cite{Tran8,n2}. We will briefly discuss the experimental range of values of the parameters 
%constants 
of our theoretical setup in a potential superconducting qubit implementation.
\par The decoherence time of a superconducting qubit  lies between $50 - 100 \mu s$ \cite{sup}. So, for our system to be useful for a superconducting qubit set-up, we must work in this time-span. We have taken our dimensionless time as $t=\delta\tilde{t}$. Let us take $\tilde{t}_{max}=50\mu s$. For $t = 0.1$, $\delta_{min} = 2 MHz$. Now, the dimensionless quantity, \(T_X\), is given by
\begin{equation}
T_{X}=\frac{K_B \tilde{T}_{X}}{\hbar \delta}.
\end{equation}  
If we choose \(T_{B}=0.2\), we would have   \(\tilde{T}_{B}|_{min}=3.04 nK\). For smaller values of $\tilde{t}$, carefully choosing $\delta$ can help in controlling the heat flow of systems with temperatures of $\mathcal{O}(mK)$ to $\mathcal{O}(K)$.
%\textcolor{red}{ager sentenceTa bojha gyalo na!!}
\par If we restrict \(\tilde{t}\) to \(\mathcal{O}(ns)\), $\delta$ will be in the $GHz$ range, and temperature around $mK$. %\textcolor{red}{ager sentenceTa bojha gyalo na!! eTa ki tar ager sentence-er ekTa example. prathamata, tahale separate para habe na. Secondly, \(\tilde{t}\) kamale, \(\delta\) baRbe, ebang tahale, \(T_X\) kambe. kintu ekhane, \(T_X\) baRchhe kibhabe? ei niye ekTa discusion hayechhilo. kintu bhule gechhi. ei goTa paraTai, ebang agerTar sesh lineTa, eksathhe bose likhe newa jabe.} 
In case of a superconducting qubit, gate lengths are of $\mathcal{O}(10 - 100ns)$, carrying out $10^{4}$ operations per coherence time \cite{sup}. Therefore, the three TLS-bath QTT set-up can be useful for controlling the heat flow of such a system. A QTT using a superconducting transmon qubit is proposed in \cite{Tran8}, where $\delta = \mathcal{O}(GHz)$.
%\par \textcolor{blue}{Even though the above discussion is based on a superconducting qubit, our set-up can be used in other systems as well. Depending on the physical qubit (superconducting, trapped-ion, solid-state spin etc.), the value of $\delta$ can be adjusted to control the heat current at the desired temperature/time range.}
%\par 

\section{Conclusion}
\label{arati-mukhopadhyay}
We were interested to find whether the three two-level systems setup, which behaves like a thermal transistor in the steady state regime, 
%has 
%some superior 
%an
%advantage 
%when operating 
can also operate 
in the transient regime, and whether there is any advantage to do so. From our analysis of several paradigmatic families of three-qubit states, including the Greenberger-Horne-Zeilinger 
and W states, as initial states of  time evolution of the  device of three qubits and three baths,
%observations by accumulating $350$ randomly chosen initial states we have got some transient thermal transistors having different amplifications and operating regions. From the study of the time evolution of some paradigmatic initial states like $GHZ$, $W$, $\ket{000}$ we have got 
we find transient thermal transistors having an efficiency better than the steady state one but having a smaller or equal operating region,
%and from the initial states $\ket{001}$ and $\ket{011}$ we can get two 
as well as
necessarily transient transistors having better amplification capability as well as a larger span of operating region.

\section{Acknowledgement}
We acknowledge support from the Department of Science and Technology,
Government of India through the QuEST grant (grant number DST/ICPS/QUST/
Theme3/2019/120).

%\newpage
\appendix
\section{Paradigmatic initial states: $W$ and $\ket{000}$ states}
\label{W,000}
\begin{figure}[h]
\includegraphics[width=\linewidth]{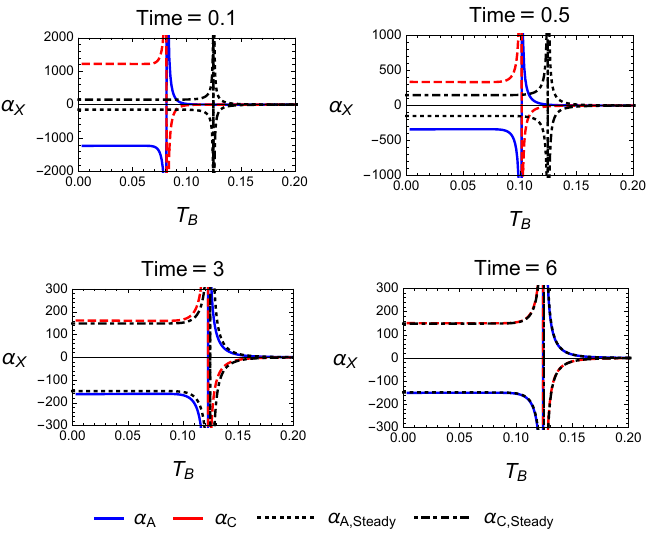}
\caption{Transient three-qubit quantum thermal transistor with the time evolution beginning in the 
%Werner 
\(W\)
state. The panels exhibit the nature of amplification factors with the change of $T_B$ for different dimensionless times. The initial state is taken as the $W$ state. The axes represent dimensionless parameters in all panels.}
\label{fig:w}
\end{figure}

The nature of variation of amplification factor for the $W$ state (Fig. \ref{fig:w}) follows a trend similar to that of the $GHZ$ state (Sec. \ref{sec:ghz}). For small time ($t=0.1$), the setup behaves as a good transistor in the region $0<T_B<0.8$, and for longer times the values of the amplification factor reduce and the operating region of the transistor in $T_B$ approaches the steady state region and reaches the steady state value at $t \approx 6$. The main difference with the GHZ state is that the amplification factor is never less than the steady state and thus the transient transistor is always better than a steady state transistor for the same discrete values of time, but for a smaller region of $T_B$. 
%These two states are two important 3 qubit entangled states and we can clearly categorise these as a better transient transistor for a small region of $T_B$.
\par
\begin{figure}[h]
\includegraphics[width=\linewidth]{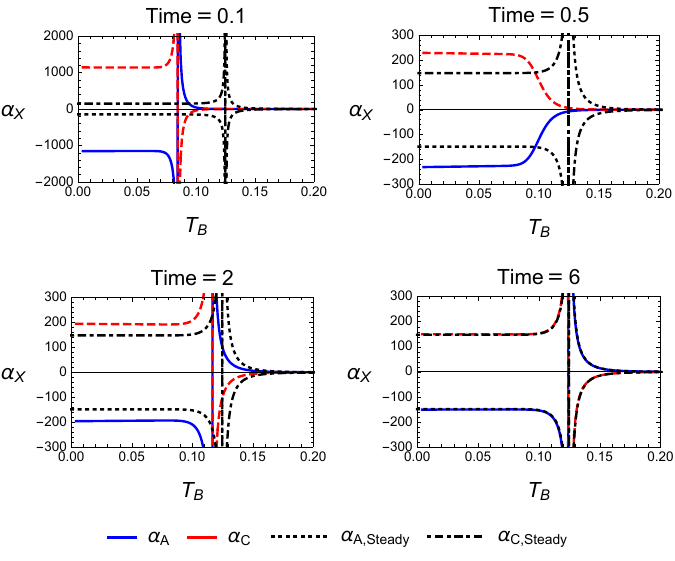}
\caption{Transient three-qubit quantum thermal transistor with the time evolution beginning in the $\ket{000}$ state. The panels exhibit the nature of amplification factors with the change of $T_B$ for different dimensionless times. The initial state is taken as the $\ket{000}$ state. The axes represent dimensionless parameters in all panels.}
\label{fig:000}
\end{figure}

 In case the initial state is chosen as 
%of 
the product state $\ket{000}$ (Fig. \ref{fig:000}), the qualitative nature of the amplification factor is the same as that of the $W$ state for very small time ($t \approx 0.2$), and then it takes the nature similar to that of the $GHZ$ state. This means that the amplification factor reduces and is almost zero for $t=0.8$ and it again has a better amplification after $t=0.9$ and, at $t \approx 6$, it reaches the steady state. As in the previous two cases, the operating region of the transient transistor is bounded for a small region of $T_B$. Therefore, for this state, too, the higher amplifications for smaller $T_{B}$ values is an interesting feature for going over to the transient regime.
\section{Initial states for obtaining necessarily transient quantum transistors} 
\label{values}
Some states that show necessarily transient behavior are shown in Fig. \ref{fig:nt} in the main text. The exact details of these states are as follows. \par
The state corresponding to the plots in Figs. \ref{fig:nt}(a) and \ref{fig:nt}(a') is $\ket{\psi^{\prime}_{GHZ}}$. It has the form as in Eq.~(\ref{eq:ghz}) with
\begin{eqnarray}
&&a=(-0.5446, -0.5546),\nonumber\\
&&b=(-0.6614, -0.1799),\nonumber\\
&&c=(0.4376, 0.4659),\nonumber\\
&&d=(-2.2000, 0.3749),\nonumber\\
&&a_1=(-1.0505, 0.2633),\nonumber\\
&&b_1=(-0.4266, -0.4274),\nonumber\\
&&c_1=(-0.9067, 0.9039),\nonumber\\
&&d_1=(0.1572, 2.3707).
\label{val:ghz}
\end{eqnarray}
All complex numbers are written as (real part, imaginary part) and the numbers have been rounded off  to four decimal places. \par
The plots in Figs. \ref{fig:nt} ($b$) and ($b'$) are obtained from an initial biseparable state, $\ket{\psi^{\prime}_{AB:C}}$, that has the form in Eq. (\ref{eq:12:3}) with
\begin{eqnarray}
&&a=(-0.2506 , -1.2750),\nonumber\\
&&b=(0.4573 , 0.0094),\nonumber\\
&&c=(1.1436 , 0.5672),\nonumber\\
&&d=(-0.9806 , 1.2475),\nonumber\\
&&a_1=( -0.7718 , 0.4604),\nonumber\\
&&b_1=(0.2562 , -0.3517).\;\;\;\;\;\;
\label{Val:12:3}
\end{eqnarray}
\par 
The $W$-class  state, $\ket{\psi^{\prime}_{W}}$, which results in the dynamics corresponding to the plots in Figs.~\ref{fig:nt} ($c$) and ($c'$), has the form in Eq. (\ref{eq:W}) with
\begin{eqnarray}
&&a=(-0.6549 , -1.5778),\nonumber\\
&&b=(0.1125 , -0.4555),\nonumber\\
&&c=(0.8575 , -0.4032),\nonumber\\
&&d=(-0.5980 , -1.0251).\;\;\;\;
\label{val:W}
\end{eqnarray}
\par
Finally, Fig. \ref{fig:nt} ($d$) and ($d'$) correspond to the product initial state, $\ket{\psi^{\prime}_{Product}}$, that has the form as in Eq. (\ref{eq:Product}) with \begin{eqnarray}
&&a=(0.7938, -0.4108),\nonumber\\
&&b=(1.6511, 0.8510),\nonumber\\
&&a_1=(-0.5692, 1.3391),\nonumber\\
&&b_1=(-0.5305, -0.3410),\nonumber\\
&&a_2=(-2.4324, -1.0312),\nonumber\\
&&b_2=(-1.1394, -0.7807).
\label{val:product}
\end{eqnarray}

\section{Behavior at intermediate times}
\label{extra}
\begin{figure}[h]
\includegraphics[width=\linewidth]{
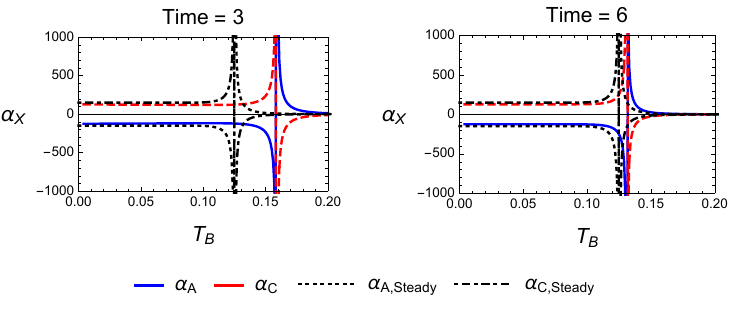}
\caption{Nature of amplification factors with the change of $T_B$ for different dimensionless times. The initial state is taken as $\ket{001}$. Both  axes represent dimensionless parameters in all panels. For the same cases at $Time=0.1$ and $Time=10$, see Fig.~\ref{fig:001}.}
\label{fig:001_app}
\end{figure}

\begin{figure}[h]
\includegraphics[width=\linewidth]{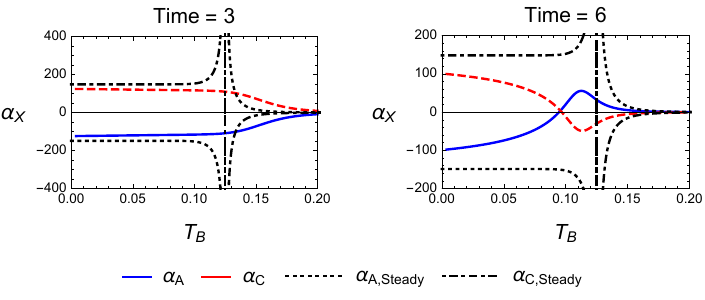}
\caption{Nature of amplification factors with the change of $T_B$ for different dimensionless times. The initial state is taken as $\ket{011}$. Both  axes represent dimensionless parameters in all panels. For the same cases at $Time=0.1$ and $Time=10$, see Fig.~\ref{fig:011}.}
\label{fig:011_app}
\end{figure}

The plots of amplification factors for different intermediate times %($t$) 
for the initial states, $\ket{001}$ and $\ket{011}$, are given in 
%below. See 
Figs.~\ref{fig:001_app} and \ref{fig:011_app}. Please refer to Sec. \ref{sec:ghz} for the details.
\par
\begin{figure}[h]
\includegraphics[width=\linewidth]{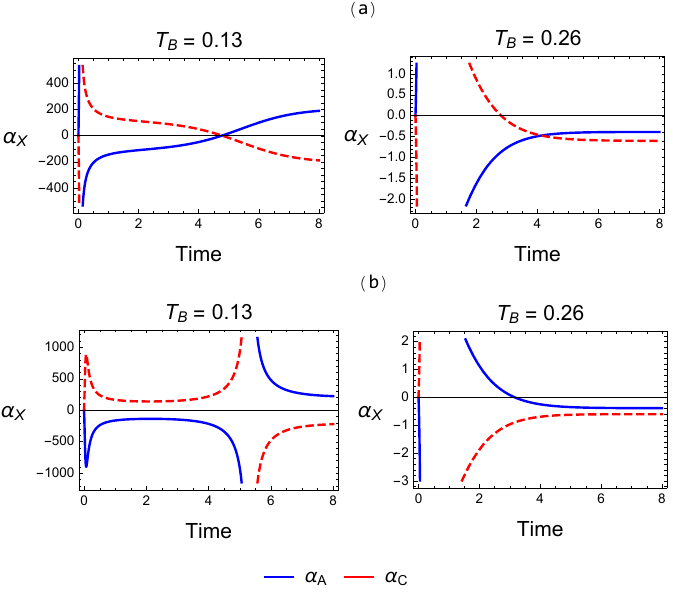}
\caption{Time evolution of amplification factor for different $T_B$ for initial states taken as $(a) \;\;\ket{\psi^{\prime}_{GHZ}}$ and $(b)\;\;\ket{\psi^{\prime}_{W}}$. All the axes represent dimensionless parameters. Plots for $T_B = 0.05$ and $T_B = 0.36$ are in Fig.~\ref{fig:time}.}
\label{fig:time_app}
\end{figure}

The plots of time variation of amplification factors for different $T_{B}$ for the initial states $\ket{\psi'_{GHZ}}$ and $\ket{\psi'_{W}}$ are given in Fig. \ref{fig:time_app}. Please refer to Sec. \ref{sec:time_vary} for the  details.

\end{document}